\documentclass{article}
\usepackage{amsmath}
\usepackage{booktabs}
    \PassOptionsToPackage{numbers, compress}{natbib}

\usepackage[preprint]{neurips_2025}
\usepackage{graphicx}
\usepackage{gensymb}
\usepackage[toc,title]{appendix}

\usepackage[utf8]{inputenc} 
\usepackage[T1]{fontenc}    
\usepackage{hyperref}       
\usepackage{url}            
\usepackage{booktabs}       
\usepackage{amsfonts}       
\usepackage{nicefrac}       
\usepackage{microtype}      
\usepackage{xcolor}         
\usepackage[skins,breakable]{tcolorbox}
\usepackage{listings}

\title{OpenFOAMGPT 2.0: end-to-end, trustworthy automation for computational fluid dynamics}

\author{
  Jingsen Feng\\
   Department of Engineering\\
  University of Exeter \\
  \And
  Ran Xu\\
  Cluster of Excellence SimTech \\
  University of Stuttgart \\
  \And
  Xu Chu \\
  Department of Engineering\\
  University of Exeter \\
  \texttt{x.chu@exeter.ac.uk} \\
}

\begin{document}

\maketitle

\begin{abstract}

We propose the first multi-agent framework for computational fluid dynamics that enables fully automated, end-to-end simulations directly from natural-language queries. The approach integrates four specialized agents—Pre-processing, Prompt Generation, OpenFOAMGPT (simulator), and Post-processing—decomposing complex computational fluid dynamics  workflows into collaborative components powered by large language models. Extensive validation through diverse case studies, including Poiseuille flows, single- and multi-phase porous media flows, and aerodynamic analyses, demonstrates 100\% success and reproducibility rates across over 450 simulations. Rigorous trustworthiness verification confirms that properly designed multi-agent systems can achieve the reliability standards necessary for zero-tolerance scientific computing applications while significantly lowering entry barriers. The framework establishes a foundation for conversation-driven simulation workflows in computational science, potentially accelerating discovery and innovation through more accessible tools for complex numerical simulations. Results reveal that multi-agent architectures, when properly specialized and orchestrated, can effectively handle the stringent requirements of computational physics while maintaining the intuitive interface of natural language interaction.
\end{abstract}

\section{Introduction}
The rapid advancement of Large Language Models (LLMs) has catalyzed a transformative era in artificial intelligence, propelling multi-agent systems into a new phase of robust development and application \citep{liu2024deepseek,huang2024crispr,buehler2024mechgpt,ghafarollahi2024rapid}. These multi-agent frameworks, which orchestrate multiple specialized AI entities toward solving complex tasks, have demonstrated remarkable versatility across disciplines including sociology, economics, medicine, pharmaceuticals, computer science, mathematics, and engineering \citep{park2023generative,dong2024self,zhang2023building,tang2023medagents,weiss2024rethinking,boiko2023emergent,chen2025mdteamgpt,li2024survey,yu2024fincon,kim2024mdagents,tang2024worldcoder}, and are even capable of conducting scientific research independently \citep{jansen2024discoveryworld,lu2024ai}. Indeed, multi-agent architectures are increasingly recognized as an effective approach for maximizing the capabilities of LLMs through specialization, collaboration, and iterative refinement processes.

Computational Fluid Dynamics (CFD), a branch of fluid mechanics employing numerical methods and algorithms to analyze and solve problems involving fluid flows \cite{wang2022spatial,Beck.2019,Duraisamy.2019,vinuesa2022enhancing,Wang.2021,wu2018physics}, represents a domain with particularly widespread applications \cite{bleeker2025neuralcfd}—spanning aerospace engineering \cite{wang2024optimized}, weather prediction \citep{lam2023learning}, biomedical flows \citep{liu2021simulation,wang2024investigation}, and Geo-Engineering \citep{yang2024data,xie2021self}. The computational complexity, domain expertise requirements, and labor-intensive workflows associated with CFD simulations have historically limited its accessibility and practical deployment despite its critical importance across scientific and engineering disciplines.

The integration of LLMs with CFD presents a compelling opportunity to address these limitations \citep{dong2025fine,du2024large,xu2024trainingmicrorobotsswimlarge}. However, this integration faces significant challenges: CFD simulations demand precise numerical specifications, complex geometry handling, sophisticated physics modeling, and specialized result interpretation. In this paper, we build upon OpenFOAMGPT system \citep{pandey2025openfoamgpt,wang2025status} to present a comprehensive end-to-end multi-agent framework specifically designed for CFD applications. Through careful agent specialization and orchestration, we achieve truly conversational simulation capabilities, where natural language queries are automatically transformed into complete, executable CFD simulations with appropriate visualization and analysis of results, while maintaining exceptional levels of trustworthiness and reproducibility. This system represents a significant advancement in human-computer interaction for scientific computing, effectively democratizing access to sophisticated fluid dynamics simulations while maintaining numerical rigor and solution quality.

By enabling end-to-end "conversation-to-simulation" workflows, our multi-agent framework reduces the technical barriers to CFD utilization while simultaneously increasing productivity for domain experts. Experimental validation across diverse flow scenarios demonstrates the system's robustness, accuracy, and practical utility for both academic and industrial applications.

\section{Multi-Agent LLM framework for CFD}
\label{gen_inst}
\begin{figure}
    \centering
    \includegraphics[width=1\linewidth]{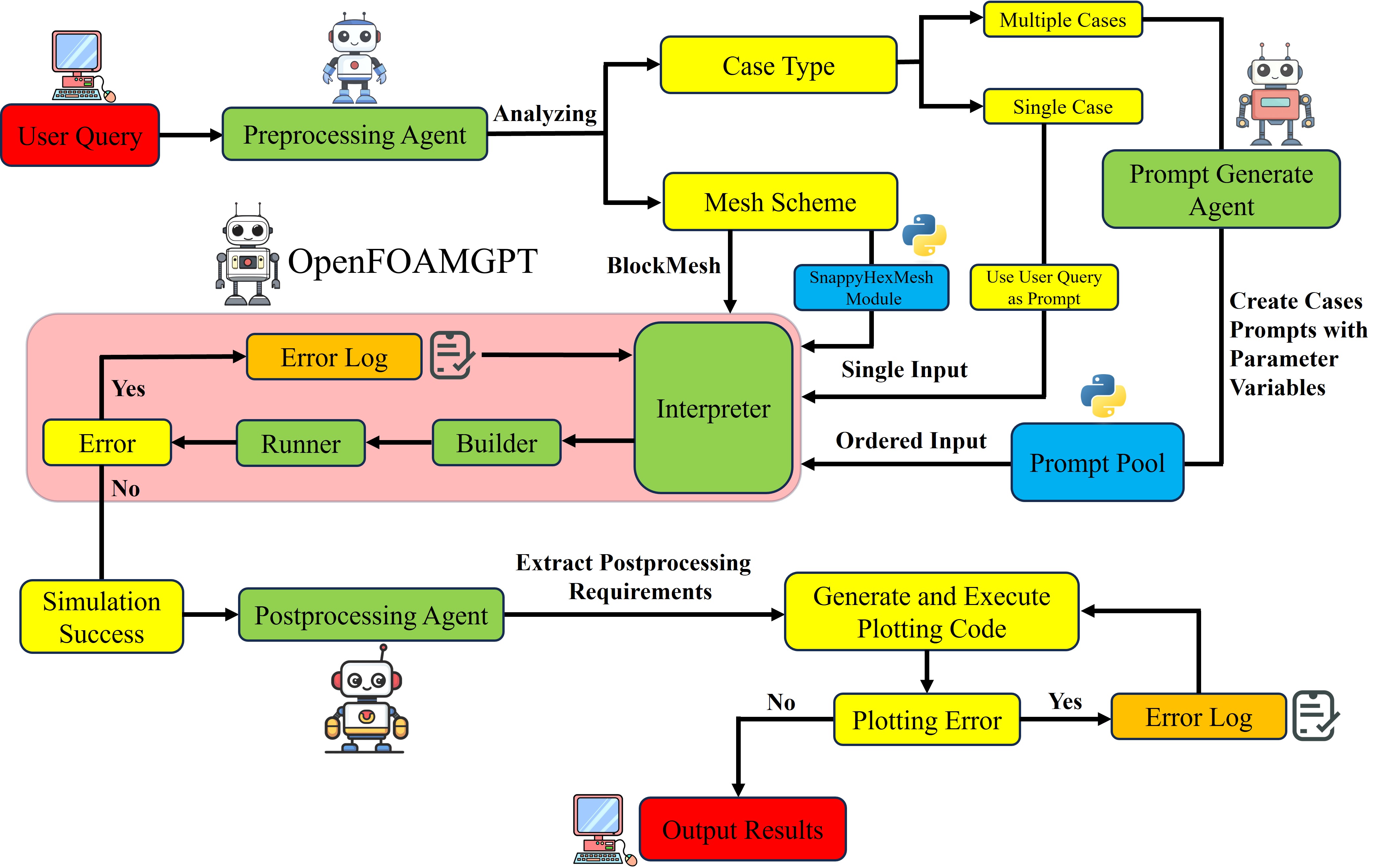}
    \caption{The design of multi-Agent framework for CFD. \textbf{Red} components represent user interfaces for input queries and output results. \textbf{Green} components indicate LLM-based intelligent agents. \textbf{Blue} components show built-in code modules working alongside agents to implement specific functionalities. \textbf{Yellow} components denote information and states provided during workflow execution. \textbf{Orange} components represent text-based output logs.}
    \label{fig:agent}
\end{figure}

The proposed multi-agent framework for CFD automation, as illustrated in Figure \ref{fig:agent}, comprises four specialized intelligent agents that collaboratively orchestrate end-to-end CFD workflows from pre-processing, simulation to post-processing without any human intervention. This framework represents a significant advancement in automated CFD simulation by integrating LLMs:

\paragraph{Pre-processing Agent:}This agent harnesses LLMs' advanced semantic understanding capabilities to comprehensively analyze user queries. 

In CFD simulations, mesh generation is a fundamental step that significantly impact the accuracy and comprehensiveness of numerical results. For mesh generation, which discretizes the computational domain into small control volumes, OpenFOAM \citep{weller1998tensorial} provides two primary utilities: \texttt{blockMesh} for structured hexahedral meshes in simple geometries, and \texttt{snappyHexMesh} for unstructured meshes that can accurately capture flow fields around complex geometries, enabling detailed resolution of boundary layers, wake regions, and other critical flow features in the vicinity of intricate geometric configurations. The quality and resolution of these spatial discretizations directly impact the accuracy of the numerical solutions and the stability of the computational process. 

Additionally, CFD studies typically require multiple simulation cases with controlled parameter variations to investigate their effects on the flow physics, rather than relying on single-case results. This parametric study approach is essential for understanding the sensitivity of flow behavior to various input parameters, establishing correlations between design variables and performance metrics, and validating numerical models across different operating conditions.

Recognizing these essential requirements, this agent intelligently determines critical simulation parameters through two primary decision paths: (1) simulation scope classification (single-case vs. multi-case parametric studies) and (2) mesh generation strategy selection (\texttt{blockMesh} for simple geometries vs. \texttt{snappyHexMesh} for complex geometries). For parametric studies, the agent automatically identifies key variables (e.g., mesh resolution, physical properties, boundary conditions). This agent defaults to single-case simulations with \texttt{blockMesh}-generated grids for computational efficiency in simple geometries.
The agent employs contextual reasoning to resolve ambiguities in problem specifications, ensuring robust parameter formalization before downstream processing. By automating these critical pre-processing decisions, the agent significantly reduces the manual effort required in traditional CFD workflows while ensuring the generation of high-quality computational setups suitable for accurate flow simulations.

\paragraph{Prompt Generation Agent:}This agent implements an adaptive prompt engineering pipeline that dynamically constructs simulation instructions tailored to each specific case requirement. This agent demonstrates remarkable capability in precisely identifying all parameters requiring adjustment across the simulation space, even in complex scenarios involving multiple interacting variables—a capability we will demonstrate in Section \ref{subsec:porous_media}.

For single-case scenarios, it directly employs user query as input prompt. In multi-case scenarios, it systematically decomposes parametric variables (e.g., mesh resolution, physical properties like viscosity or density, boundary conditions) to generate a structured set of case-specific prompts. This approach of generating dedicated prompts for each parametric variation was deliberately chosen over alternative methods. 

A seemingly straightforward alternative would be to replicate configuration files from an initial successful simulation and subsequently modify specific parameters for derivative cases. However, this approach presents significant challenges in the context of OpenFOAM, whose configuration files adhere to extremely strict syntactical requirements and structural hierarchies. Such an approach would necessitate an additional specialized agent to guide the LLM in making precise modifications to the correct parameter locations across multiple interconnected files. When dealing with multi-variable modifications, this method becomes highly inefficient and prone to errors, as even minor syntax inconsistencies can cause simulation failures that are difficult to diagnose and correct automatically. By contrast, our approach of generating fresh, comprehensive prompts for each simulation case offers superior efficiency, robustness, and adaptability. These prompts are systematically organized in a Prompt Pool repository (Figure \ref{fig:agent}), which serves as a dynamic knowledge base for orchestrating sequential parametric studies. This methodology ensures that each simulation case receives a complete, coherent set of instructions, eliminating the risks associated with incremental file modifications while maintaining a high level of automation across diverse simulation requirements.

\paragraph{OpenFOAMGPT:} This agent functions as the core simulation engine\citep{pandey2025openfoamgpt,wang2025status}, executing a robust self-correcting simulation loop through three sophisticated modules that work in concert to ensure successful CFD simulations:

1. Configuration Generation: Synthesizes comprehensive OpenFOAM case configurations by intelligently mapping task-specific requirements to appropriate numerical schemes and solver parameters. This module serves as the critical foundation for simulation accuracy, as the correctness and completeness of configuration files directly determine whether OpenFOAM can properly execute the simulation. The system employs a carefully constructed system prompt that constrains the LLM's output format and structure. Each case-specific prompt is processed alongside this system prompt, enabling the LLM to understand specific simulation requirements while generating syntactically precise configuration files. Given the extremely stringent formatting requirements of OpenFOAM configuration files, where even minor syntax errors can cause simulation failure, we deliberately avoid using the LLM's "thinking" capabilities and instead set the temperature parameter to 0. This design choice ensures consistent, deterministic outputs, eliminating the variability that could introduce subtle errors in dictionary structures, parameter specifications, or boundary condition definitions. 

2. Automated Execution Management: Orchestrates simulation workflows in executed OpenFOAM environment. While this module does not directly involve LLM inference, it represents an essential component in the automation pipeline. The execution environment utilizes the OpenFOAM v2406 Docker container, providing a standardized, reproducible runtime environment independent of the host system configuration. This module automatically executes the "\texttt{Allrun}" script generated in the previous step, monitors the simulation progress, and captures all console output and log files for potential error analysis.


3. Error-Driven Iterative Refinement: Despite the precision-focused approach in configuration generation, LLM-generated configuration files may occasionally contain errors or omit necessary components that lead to simulation failures. In such cases, this module captures detailed error logs and contextual information about the failure, structuring this feedback for the LLM. The error information is then fed back to the LLM along with the original case prompt and system constraints, enabling intelligent refinement of the configuration files. This creates a closed-loop learning process where the system iteratively improves configurations based on specific error feedback until achieving successful simulation. This self-correcting capability significantly enhances the system's robustness and success rate across diverse simulation requirements.

\paragraph{Post-processing Agent:}This agent extracts visualization and analysis requirements from original user queries, functioning as the final component in the automated CFD workflow. This agent operates on the simulation outputs systematically stored by OpenFOAM in the Post-processing directory—a structured repository containing field data, probe measurements, and derived quantities that were specified during the configuration generation phase by OpenFOAMGPT based on the requirements interpreted from each case's prompt. Upon simulation completion, this agent automatically accesses and parses these output files, intelligently matching the available data with the visualization and analysis requirements originally specified in the user query. It will generate optimized Python scripts leveraging scientific computing libraries (NumPy, Matplotlib) for data processing and visualization, while also providing Paraview VTK files for visualization analysis. Its capabilities include automated generation of publication-quality visualizations adapted to the specific simulation context—such as contour plots of pressure and velocity fields, streamlines and vector plots to illustrate flow patterns, and integrated quantities such as forces and moments.
For parametric studies involving multiple simulation cases,  the agent automatically generates comparative visualizations that highlight trends and relationships between input parameters and output results—such as parameter sensitivity plots. Through this integrated approach to Post-processing, the agent transforms raw simulation data into actionable engineering insights, completing the automated workflow from problem specification to results interpretation without requiring manual intervention at any stage.

\section{Experiments}
\label{headings}

In this section, we present a comprehensive evaluation of the proposed multi-agent, end-to-end workflow through five representative case studies that span a diverse range of CFD applications: (i) single-phase Poiseuille flow, (ii) multi-phase Poiseuille flow, (iii) single-phase flow in porous media, (iv) multi-phase flow in porous media, and (v) aerodynamics of a motorbike. These cases were deliberately selected to assess the framework's capability across varying levels of physical complexity, from fundamental channel flows to more complex scenarios. All intelligent agents within the framework are powered by the Claude-3.7-Sonnet, which provides the foundation for natural language understanding and generation capabilities. Throughout our experimental validation, the framework demonstrated remarkable robustness and reliability. When provided with well-formulated user queries and operating under the constraints of our carefully designed system prompts, the framework achieved a 100\% success rate across both single-case simulations and parametric multi-case studies. Importantly, all simulation results maintained complete reproducibility across multiple execution instances, indicating the deterministic nature of the workflow.

\subsection{Single-phase Poiseuille flow}

Single-phase Poiseuille flow represents a canonical case in fluid mechanics that describes the laminar flow of a viscous fluid through a straight channel under a pressure gradient. This flow configuration is particularly valuable as a validation case due to its well-established analytical solution, making it an ideal benchmark for verifying numerical methods. 

Our multi-agent workflow autonomously processed the user query through its complete pipeline. The Pre-processing Agent correctly identified this as a single-case simulation requiring \texttt{blockMesh} for the simple rectangular geometry. The Prompt Generation Agent transformed the user query into a structured simulation directive, which was then passed to OpenFOAMGPT. OpenFOAMGPT successfully generated all necessary configuration files, including appropriate boundary conditions for the pressure-driven channel flow, correct specification of the fluid properties, and proper numerical schemes for the \texttt{icoFoam} solver.

Once the simulation completed, the Post-processing Agent automatically extracted the velocity profile data from the sampling location specified in the prompt. Notably, without any additional guidance, the agent recognized the need for analytical comparison and implemented the appropriate Poiseuille flow equation  $u(y) = -\Delta Py(H-y)/{2\mu L}$, where $\Delta P$ is the pressure difference, $\mu$ is the dynamic viscosity, $L$ is the channel length, and $H$ is the channel height. The agent then generated a publication-quality plot comparing the numerical results with the analytical solution, demonstrating excellent agreement between the two and confirming the accuracy of the simulation (Figure \ref{singlemulti}(a)).

\subsection{Multi-phase Poiseuille flow}
Multi-phase Poiseuille flow represents a validation case for immiscible, incompressible multi-phase flows. This configuration features a stratified flow arrangement where a non-wetting phase flows between two layers of a wetting phase adjacent to the channel walls. This arrangement creates distinct interfaces between the phases and introduces complex momentum transfer mechanisms that must be accurately resolved. The analytical solution for this stratified flow \citep{yiotis2007lattice} provides a rigorous reference for validating numerical implementations. 

\begin{figure}
    \centering
    \includegraphics[width=0.9\linewidth]{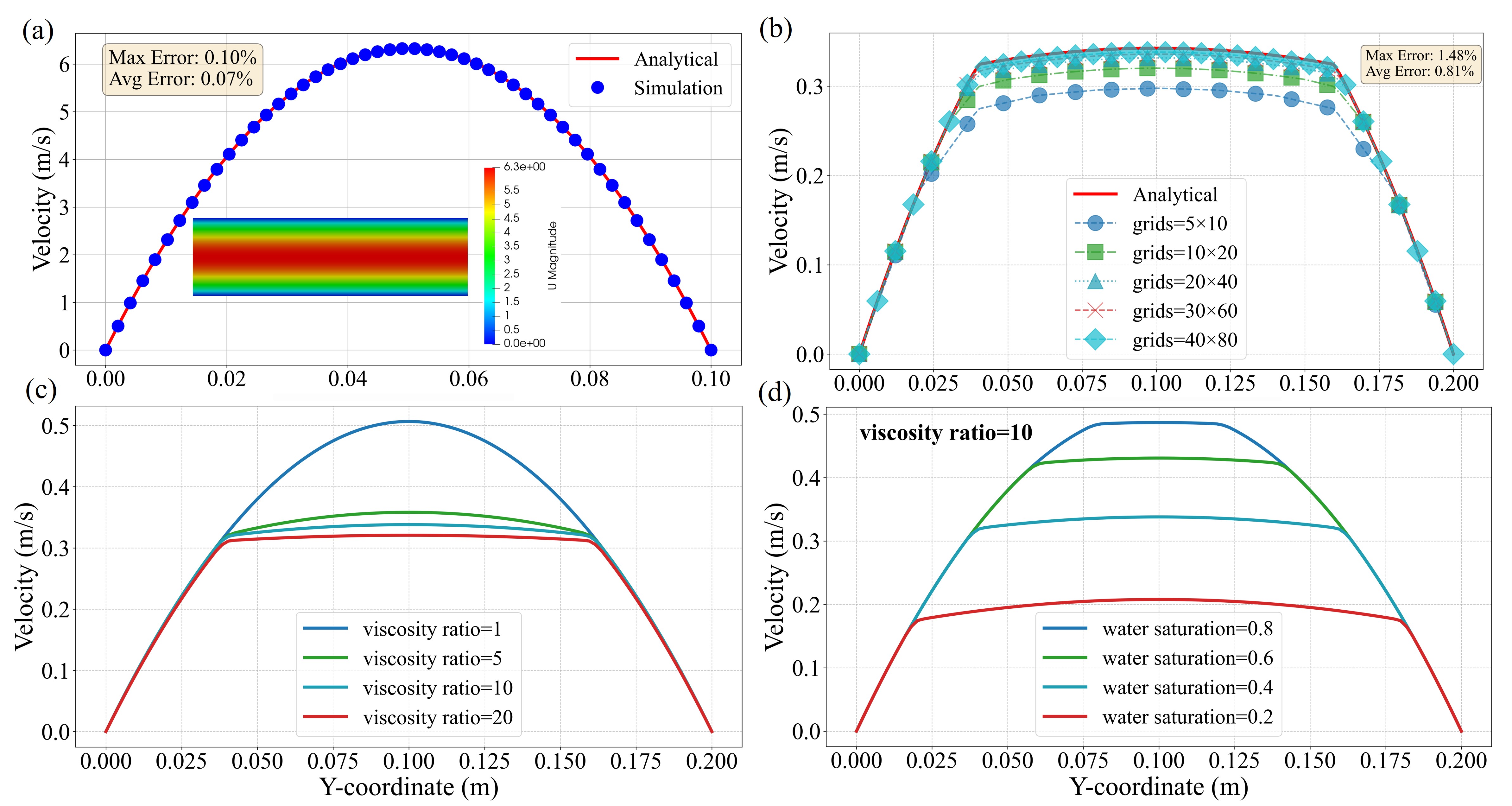}
    \caption{Validation and parametric studies of Poiseuille flow simulations. (a) Comparison between numerical results and analytical solution for single-phase Poiseuille flow; (b) Grid independence study for multi-phase Poiseuille flow; (c) Effect of viscosity ratio on velocity distribution in multi-phase Poiseuille flow; (d) Influence of wetting phase (water) saturation on velocity profiles in multi-phase Poiseuille flow. All subfigures are generated by the OpenFOAMGPT 2.0}
    \label{singlemulti}
\end{figure}

To comprehensively evaluate our multi-agent framework's capability in handling this more complex multi-phase scenario, we designed a parametric study comprising three series of continuous simulations: 
\paragraph{1. Grid Independence Study:}To establish the numerical accuracy and convergence characteristics of the solution, we conducted simulations with progressively refined mesh resolutions: $5\times10$, $10\times20$, $20\times40$, $30\times60$, and $40\times80$ cells,as shown in Figure   \ref{singlemulti}(b).
\paragraph{2. Saturation Variation Study:}To investigate the influence of phase distribution on flow behavior, we varied the wetting phase saturation (the volume fraction of the wetting phase relative to the total fluid volume) across four cases: $0.2$, $0.4$, $0.6$, and $0.8$, as shown in Figure \ref{singlemulti}(c).
\paragraph{3. Viscosity Ratio Study:} To examine the system's response to varying degrees of viscous coupling, we implemented simulations with viscosity ratios ($\mu_{nw}/\mu_w$) of $1$, $5$, $10$, and $20$, as shown in Figure \ref{singlemulti}(d).

Our multi-agent workflow correctly identified this as a multi-case simulation requiring multiphase physics. The Prompt Generation Agent successfully created individual prompts for each distinct simulation cases. OpenFOAMGPT correctly generate configuration files for all cases, and the simulations completed successfully.
The Post-processing Agent automatically generated comparative visualizations showing grid convergence, saturation, and viscosity ratio influences. The successful execution of three continuous simulations demonstrates our framework's robust capability to handle complex parametric studies without manual intervention.

\subsection{Single-phase flow in porous media}
\label{subsec:porous_media}
Fluid flow through porous media represents a common engineering phenomenon encountered in various applications ranging from groundwater flow to oil recovery. The complex geometric boundaries inherent in porous structures necessitate the use of unstructured meshes, as structured meshes generated by \texttt{blockMesh} cannot adequately represent the intricate flow paths. This case study evaluates our multi-agent workflow's capability to correctly employ \texttt{snappyHexMesh} for generating unstructured meshes around complex geometries and subsequently simulate the flow physics. 

\begin{figure}
    \centering
    \includegraphics[width=1\linewidth]{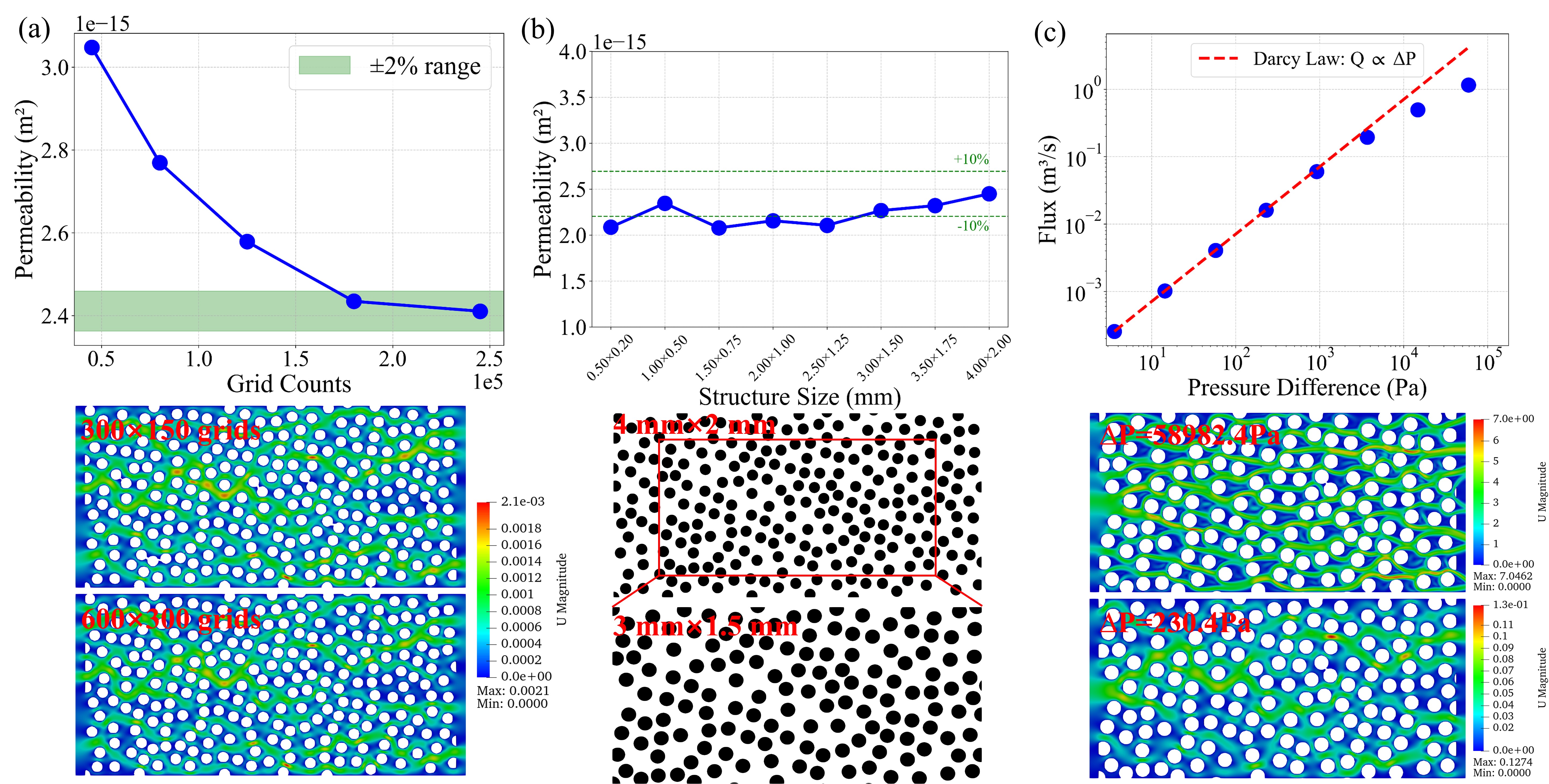}
    \caption{Results of single-phase flow in porous media simulation. (a) Grid independence study of permeability; (b) Determination of permeability REV size; (c) Relationship between flux and pressure difference. All three upper figures are generated by OpenFOAMGPT 2.0}
    \label{singleporous}
\end{figure}

We selected a 2D packed circular geometry with dimensions of $4\text{ mm} \times 2\text{ mm}$ and porosity of $0.5567$ as our test case. The workflow autonomously executed three series of simulations to characterize this porous structure:

First, a mesh independence study was conducted to ensure solution accuracy, employing resolutions of $300\times150$, $400\times200$, $500\times250$, $600\times300$, and $700\times350$ cells. Buffer regions comprising $0.025\%$ of the domain length were added to both sides to ensure proper implementation of pressure boundary conditions. Our multi-agent system correctly identified that permeability calculations with the $700\times350$ mesh differed by less than $2\%$ from those with the $600\times300$ mesh, confirming sufficient resolution with the latter, as shown in Figure \ref{singleporous}(a).

Subsequently, the framework conducted a permeability representative elementary volume (REV) 
 \citep{liu2022critical} determination study. This involved eight continuous simulations with domain sizes of $0.5\times0.25$, $1\times0.5$, $1.5\times0.75$, $2\times1$, $2.5\times1.25$, $3\times1.5$, $3.5\times1.75$, and $4\times2$ mm. To maintain consistent mesh resolution and pressure gradients across scales, the system automatically adjusted grid counts to $75\times38$, $150\times75$, $225\times113$, $300\times150$, $400\times200$, $450\times225$, $525\times263$, and $600\times300$, while scaling pressure differences proportionally from $0.06$ to $0.48$ Pa. Despite the complexity of simultaneously varying geometry dimensions, mesh counts, and pressure conditions, our workflow correctly identified all simulation parameters for each case and successfully determined the REV size as $3\times1.5$ mm, as shown in Figure \ref{singleporous}(b).

Finally, using the established REV geometry, the system executed eight continuous simulations across different pressure gradients to characterize the flow regime. The Post-processing agent automatically generated plots that clearly illustrated the transition from Darcy flow (where flow rate varies linearly with pressure gradient according to $q = -k\nabla p/\mu$ at low pressure differences to non-Darcy flow at higher gradients, where inertial effects become significant and the relationship deviates from linearity, as shown in Figure \ref{singleporous}(c).\vspace{1em} ~\\
This case study demonstrates our framework's ability to handle complex geometry preparation, unstructured mesh generation, and physics-based analysis requiring multiple interconnected parametric studies—capabilities essential for practical engineering applications beyond simple benchmark cases.

\subsection{Multi-phase flow in porous media}

\begin{figure}
    \centering
    \includegraphics[width=1\linewidth]{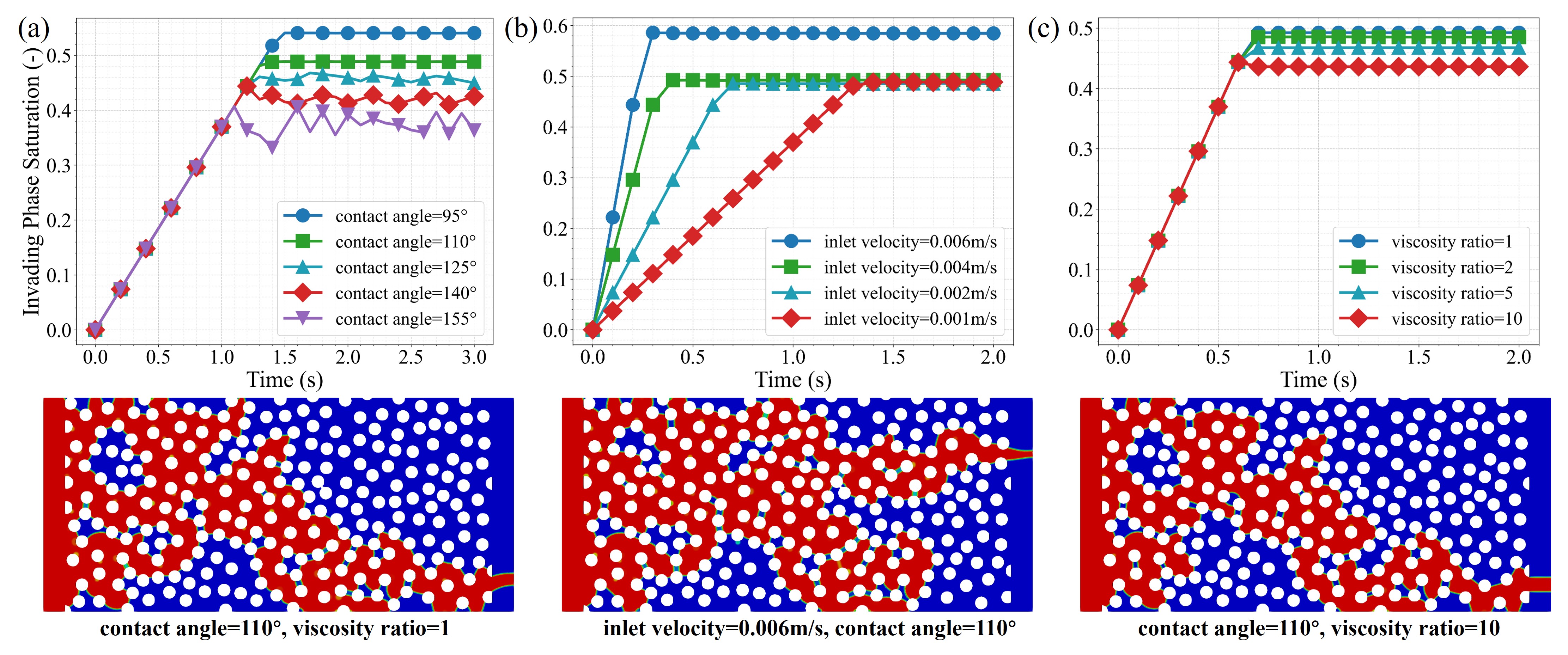}
    \caption{Displacement efficiency in drainage process of multi-phase flow through porous media. (a) Effect of contact angle variation; (b) Influence of inlet velocity; (c) Impact of viscosity ratio between displaced and invading fluids. All three upper figures are generated by OpenFOAMGPT 2.0}
    \label{displacement}
\end{figure}

Building upon the porous geometry established in Section \ref{subsec:porous_media}, we extended our investigation to multi-phase flow dynamics by simulating drainage processes—a phenomenon of significant importance in petroleum engineering and hydrogeology where a non-wetting phase displaces a wetting phase from a porous medium \citep{xu20172.5,xu2017microfluidic}. Using the $4\text{ mm} \times 2\text{ mm}$ porous structure with porosity of $0.5567$, our multi-agent workflow autonomously executed three parametric studies to investigate the influence of key parameters on drainage efficiency:
\paragraph{1. Contact Angle Study:} Five simulations with the invading non-wetting phase at a fixed inlet velocity of $0.001\text{ m/s}$ while varying the contact angle at $95^\circ$, $110^\circ$, $125^\circ$, $140^\circ$, and $155^\circ$, as shown in Figure \ref{displacement}(a).
\paragraph{2. Injection Velocity Study:}Four simulations with a fixed contact angle of $110^\circ$ and viscosity ratio of $1$ while varying the invading phase velocity at $0.001\text{ m/s}$, $0.002\text{ m/s}$, $0.004\text{ m/s}$, and $0.006\text{ m/s}$, as shown in Figure \ref{displacement}(b).
\paragraph{3. Viscosity Ratio Study:}Four simulations with a fixed $0.002\text{ m/s}$ inlet velocity and contact angle of $110^\circ$ while varying the viscosity ratio between displaced and invading fluids at $1$, $2$, $5$, and $10$, as shown in Figure \ref{displacement}(c). 

Our multi-agent workflow successfully executed all three parametric simulations, correctly handling complex geometry preparation, interface physics, and parameter variations across contact angles, injection velocities, and viscosity ratios. The system autonomously generated visualizations and identified displacement efficiency, demonstrating the framework's capability to conduct sophisticated multi-phase investigations without human intervention.

\subsection{Aerodynamics of a motorbike}

Aerodynamics of a motorbike serves as an example of turbulent flow simulations. This case study employs a provided \texttt{polyMesh} folder to demonstrate our multi-agent workflow's capability in handling pre-generated mesh geometries. The investigation characterizes a motorcycle's aerodynamic behavior across ten velocity conditions ranging from $10 $ \text{m/s} to $100 $ \text{m/s} in $10$ \text{m/s} increments.

\begin{figure}
    \centering
    \includegraphics[width=0.8\linewidth]{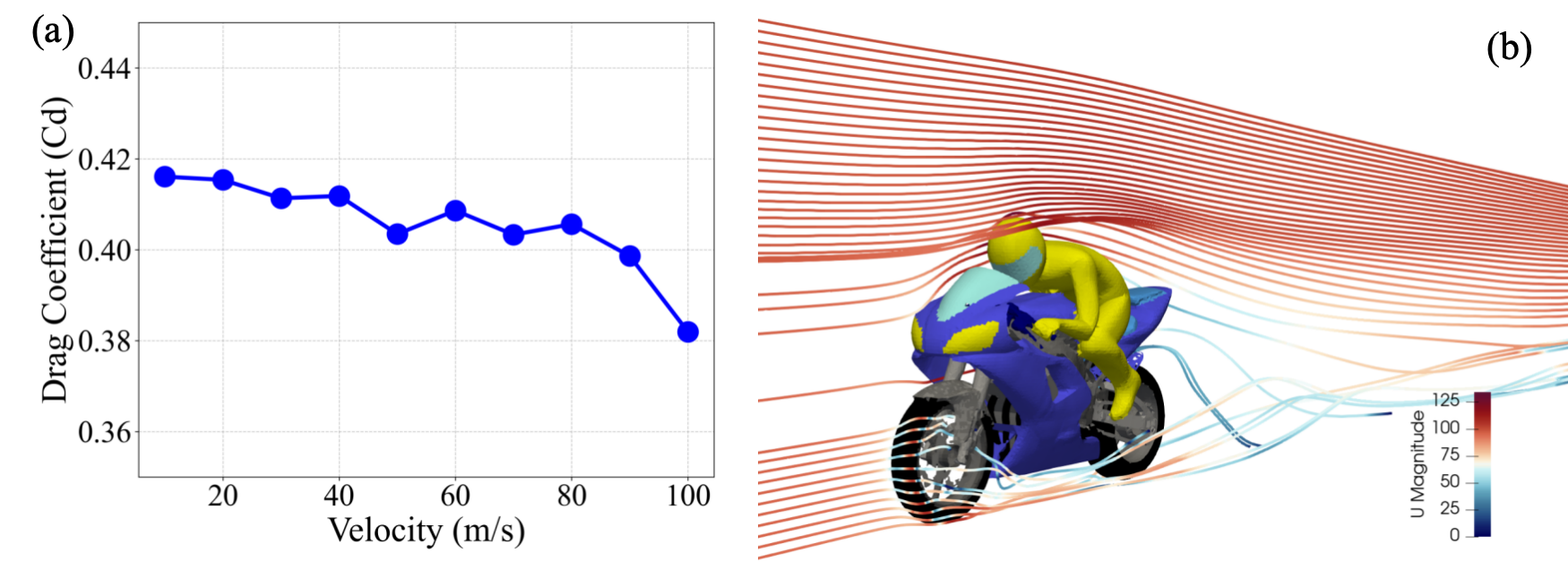}
    \caption{Results of aerodynamics of a motorbike  simulation. (a) Drag coefficient ($C_d$) in different velocity; (b) The streamline flow characteristics surrounding a motorbike operating at $100\text{m/s}$. The left figure is generated by OpenFOAMGPT 2.0}
    \label{result2}
\end{figure}

The framework conducted an aerodynamic coefficient study across the velocity. As shown in Figure \ref{result2}(a), the post-processing agent automatically extracted drag coefficients ($C_d$) from each simulation, where the drag coefficient is defined as the dimensionless quantity expressing the aerodynamic drag force normalized by dynamic pressure and reference area: $C_{\text{d}} = D/(0.5 \rho u^2 A)$, where $D$ is the drag force (N), $\rho$ is the air density (kg/m$^3$), $u$ is the flow velocity (m/s), and $A$ is the characteristic reference area (m$^2$). This analysis revealed $C_{\text{d}}$ variations across different velocities. Subsequently, the workflow analyzed streamline flow through automated visualization. Figure \ref{result2}(b) illustrates a visualization of velocity magnitude contours and streamline trajectories at $100$ \text{m/s}, highlighting key flow features including helmet wake vortices, fairing boundary layer development, and rear wheel turbulence.

This case demonstrates the capability of our multi-agent workflow to directly accept provided mesh files, validating its technical competence in maintaining simulation integrity while circumventing the geometry pre-processing phase. This characteristic substantiates the workflow's compatibility with specialized cases requiring custom meshing.

\subsection{Trustworthiness verification}

High trustworthiness is a critical requirement for CFD simulations, where even minor inconsistencies can lead to significant discrepancies in results. A well-designed CFD multi-agent system must demonstrate exceptional reproducibility to be considered reliable for practical applications. To rigorously assess this aspect of our framework, we conducted extensive verification experiments across all test cases (Table \ref{tab:trustworthiness}).

\begin{figure}
    \centering
    \includegraphics[width=0.4\linewidth]{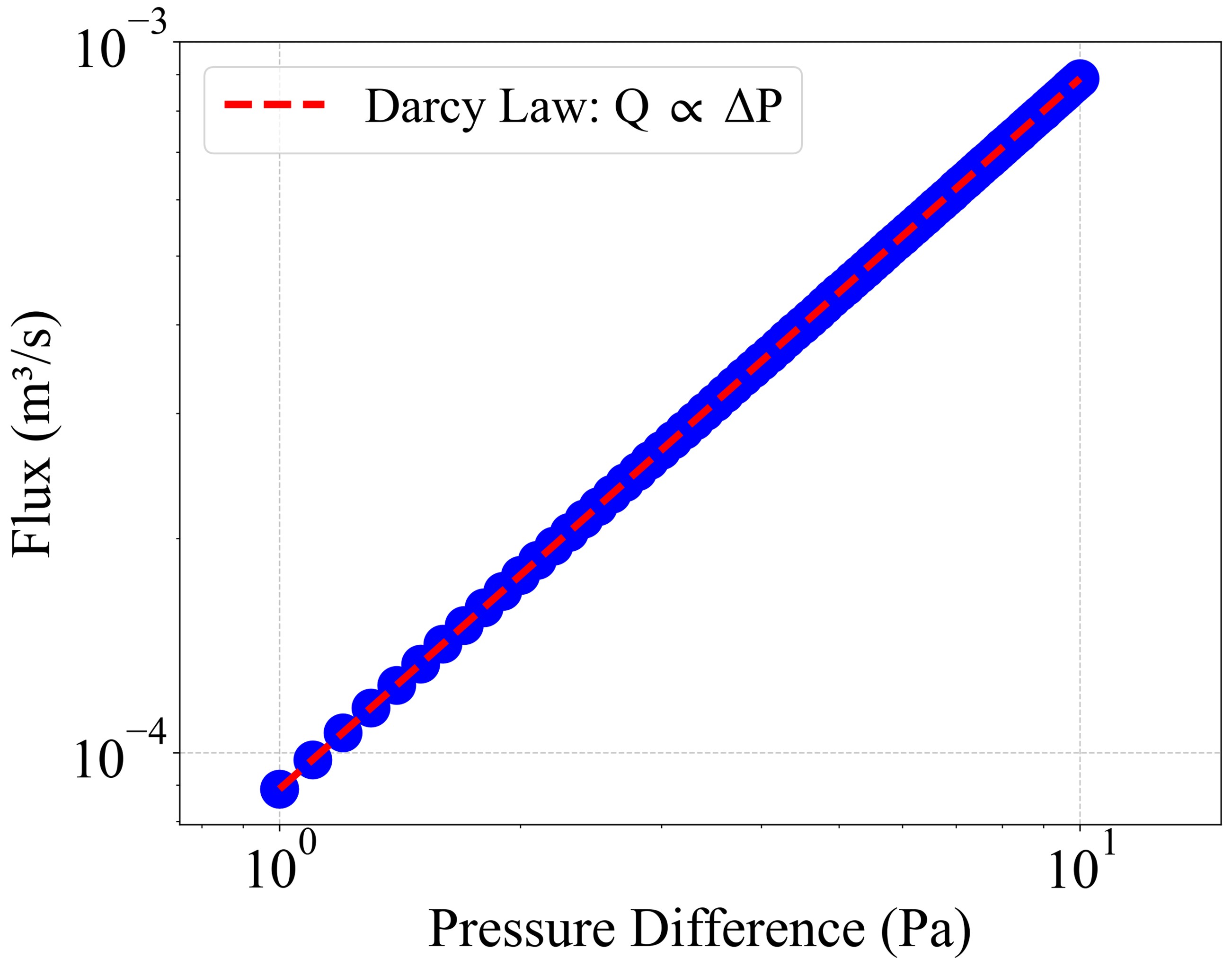}
    \caption{Results of 100 continuous simulations with the pressure difference changing uniformly from 1 Pa to 10 Pa. The figures is generated by OpenFOAMGPT 2.0}
    \label{100}
\end{figure}

\begin{table}[htbp]
\setlength{\abovecaptionskip}{1pt}
\centering
\caption{Tests of trustworthiness}
\label{tab:trustworthiness}
\resizebox{\textwidth}{!}{
\renewcommand{\arraystretch}{1.5}
\fontsize{14}{14}\selectfont
\begin{tabular}{lccccc|c}
\noalign{\hrule height 1pt}
 
\textbf{Metric} & \textbf{Single-phase} & \textbf{Multi-phase} & \textbf{Single-phase} & \textbf{Multi-phase} & \textbf{Aerodyn.} & \textbf{Single-phase}\\
 & \textbf{Poiseuille flow} & \textbf{Poiseuille flow } & \textbf{porous media} & \textbf{porous media} & \textbf{motorbike} & \textbf{porous media}\\
\hline
{Repeat count} & {10} &{10} & {10} & {5} & {10} & {2}\\
{Continuous simulations} & {1} & {4} & {8} & {5} & {10} & \textcolor{red}{100} \\
{Total simulation cases} & {10} & {40} & {80} & {25} & {100} & {200} \\
{Success rate (\%)} & {100} & {100} & {100} & {100} & {100} & {100} \\
\noalign{\hrule height 1pt}
\end{tabular}
}
\end{table}
Each case study underwent multiple executions to verify consistent behavior. Particular attention was devoted to the single-phase flow in porous media case, for this scenario, we performed 10 repetitions of the 8-case continuous simulation sequence described in Section \ref{subsec:porous_media}. Additionally, we conducted two extended experiments consisting of 100 continuous simulations each: one maintaining fixed parameters and another with varied parameters(from 1Pa to 10Pa pressure difference). As demonstrated in Table \ref{tab:trustworthiness} and Figure \ref{100}, our framework achieved a 100\% reproducibility rate across all test cases—including the most demanding multi-case parametric studies. This excellent level of trustworthiness empirically confirms the robustness of our multi-agent architecture. The results are particularly significant given the zero-tolerance nature of CFD tasks, where precise numerical specifications, complex geometry handling, and strict syntax requirements make reproducibility challenging.

These findings demonstrate that through careful system design—including specialized agent roles, precise prompt engineering, and robust error-handling mechanisms—multi-agent frameworks can achieve the high trustworthiness needed for mission-critical scientific computing applications. Such reliability represents a significant advancement toward practical deployment of LLM-based systems in domains traditionally requiring extensive human expertise and oversight.

\section{Conclusion}

This paper presents a multi-agent framework that transforms natural language queries into end-to-end CFD simulation results with visualization and analysis. By integrating specialized agents for pre-processing, prompt generation, simulation execution, and post-processing, our system successfully handles diverse fluid dynamics problems with 100\% reproducibility. OpenFOAMGPT 2.0 demonstrates that properly designed LLM-based systems can achieve the reliability standards necessary for scientific computing while significantly reducing expertise barriers. Our approach enables conversation-driven, end-to-end simulation workflows that maintain numerical rigor and physical accuracy, establishing a foundation for more accessible computational tools in engineering disciplines. The proposed architectural principle can be readily extended to other computational domains that demand comparable precision, and our future work will focus on adapting them to more complex, industry-level challenges.



\bibliographystyle{unsrtnat}
\bibliography{sn-bibliography}

\end{document}